\documentclass[aps,twocolumn,superscriptaddress,showpacs]{revtex4-1}
\usepackage{graphicx}
\usepackage{amsmath}
\usepackage{amsfonts}
\usepackage{amssymb}
\usepackage[breaklinks=true,hidelinks]{hyperref}

\newcommand{\cf}{cf.\ }
\newcommand{\dd}{\mathrm{d}}
\newcommand{\realt}{\operatorname{Re}}
\newcommand{\DC}{\Delta_\mathrm{c}}
\newcommand{\ER}{E_\mathrm{R}}
\newcommand{\etac}{\eta_\mathrm{c}}
\newcommand{\etap}{\eta_\mathrm{p}}
\newcommand{\kB}{k_\mathrm{B}}
\newcommand{\omegac}{\omega_\mathrm{c}}
\newcommand{\omegap}{\omega_\mathrm{p}}
\newcommand{\omrec}{\omega_\mathrm{R}}

\begin{document}

\title{Seeding patterns for self-organization of photons and atoms}

\author{Wolfgang Niedenzu}
\affiliation{Institut f{\"u}r Theoretische Physik, Universit{\"a}t Innsbruck, Technikerstra{\ss}e~25, A-6020~Innsbruck, Austria}
\affiliation{Theoretische Physik, Universit{\"a}t des Saarlandes, Campus~E2.6, D-66123~Saarbr\"ucken, Germany}

\author{Stefan Sch\"utz}
\affiliation{Theoretische Physik, Universit{\"a}t des Saarlandes, Campus~E2.6, D-66123~Saarbr\"ucken, Germany}

\author{Hessam Habibian}
\affiliation{Theoretische Physik, Universit{\"a}t des Saarlandes, Campus~E2.6, D-66123~Saarbr\"ucken, Germany}
\affiliation{ICFO -- Institut de Ci\`encies Fot\`oniques, Mediterranean Technology Park, E-08860~Castelldefels~(Barcelona), Spain}

\author{Giovanna Morigi}
\affiliation{Theoretische Physik, Universit{\"a}t des Saarlandes, Campus~E2.6, D-66123~Saarbr\"ucken, Germany}

\author{Helmut Ritsch}
\email{Helmut.Ritsch@uibk.ac.at}
\affiliation{Institut f{\"u}r Theoretische Physik, Universit{\"a}t Innsbruck, Technikerstra{\ss}e~25, A-6020~Innsbruck, Austria} 

\begin{abstract}
When atoms scatter photons from a transverse laser into a high-finesse optical cavity, they form crystalline structures which maximize the intracavity light field and trap the atoms in the ordered array. Stable organization occurs when the laser field amplitude exceeds a certain threshold. For planar single-mode cavities there exist two equivalent possible atomic patterns, which determine the phase of the intracavity light field. Under these premises, we show that the effect of an additional laser pumping the cavity makes one pattern more favorable than the other and that it can dynamically force the system into a predetermined configuration. This is an instance of pattern formation and seeding in a nonlinear quantum-optical system. 
\end{abstract}

\date{September 18, 2013}

\maketitle

\section{Introduction}

Pattern formation is a remarkable phenomenon of nonlinear dynamics, which characterizes the physical behavior of complex systems~\cite{cross1993pattern}. Among several realizations encountered in the quantum world, one interesting example, on which we will focus in this work, is the formation of self-organized Bragg gratings by particles scattering photons in a high-finesse optical resonator~\cite{domokos2002collective,ritsch2013cold}. This behavior is due to the mechanical effects of atom--photon interactions and emerges from coherent light scattering by an atomic gas into a single mode of the electromagnetic field inside the cavity. The light scattered by a spatially homogeneous particle distribution cannot significantly excite the cavity mode due to destructive interference of the electric fields stemming from different atoms since they have random phases. Above a certain critical intensity of the transverse pump laser, however, the atomic density undergoes a transition to a periodic distribution for which the intracavity light field is maximized. Vice versa, the field gives rise to a mechanical potential, which confines the atoms in the very same pattern. In a single-mode standing-wave cavity there exist two equivalent such configurations, each corresponding to an intracavity field of the same amplitude but of opposite phase. This behavior was first predicted in~\cite{domokos2002collective} and experimentally confirmed shortly afterwards~\cite{black2003observation}. Several theoretical~\cite{asboth2005selforganization,vukics2007microscopic,niedenzu2011kinetic,oztop2012excitations,piazza2013bose} and experimental studies~\cite{black2005collective,baumann2010dicke,baumann2011exploring,arnold2012self,brennecke2013real} have analyzed various aspects of the self-organization phenomenon, including extensions to multimode resonators~\cite{gopalakrishnan2009emergent,gopalakrishnan2010atom,gopalakrishnan2011frustration,strack2011dicke,habibian2013bose}.
\par
\begin{figure}
  \includegraphics[width=0.9\columnwidth]{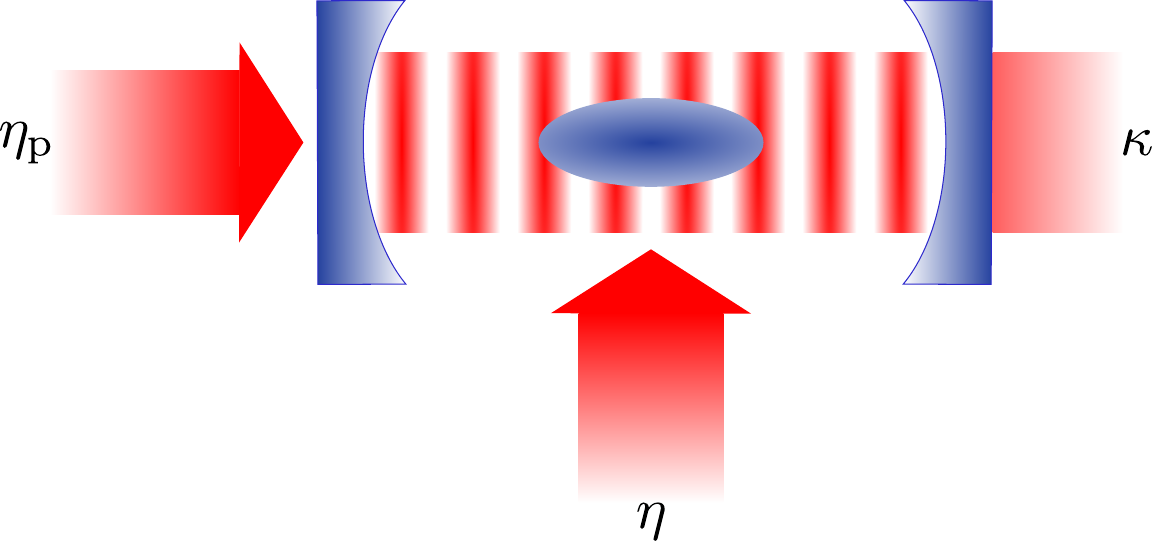}
  \centering
  \caption{(Color online) Sketch of the system. Atoms are confined inside a cavity with decay rate $\kappa$ and illuminated by a transverse laser beam of strength $\eta$. When $|\eta|$ exceeds a certain threshold value the atoms self-organize in stable Bragg gratings which maximize scattering into the cavity mode. Here, we study how these dynamics are modified when the cavity mode is additionally driven by a longitudinal laser of strength $\etap$.}\label{fig_system}
\end{figure}

Studies of dynamics of pattern formation in other systems have shown that these dynamics can be significantly modified in the presence of further pumping fields~\cite{yang2011controllable}, which can impose an auxiliary phase favoring a certain pattern. In this work we theoretically study the interplay between longitudinal and transverse laser pump fields in determining spatial ordering inside a single-mode standing-wave resonator. We consider a setup as in Fig.~\ref{fig_system} and show that, when the two lasers driving the cavity and the atoms are resonant, the laser pumping the cavity can act as a seed for the dynamics of pattern formation. We identify the conditions on the longitudinal laser for which the phase of Bragg gratings can be predetermined, and for which one can even dynamically force a Bragg grating of atoms to jump into another pattern.

This paper is organized as follows. In Sec.~\ref{sec_model} the theoretical model is introduced, from which semiclassical stochastic differential equations are derived. These equations are the basis of the numerical simulations presented in Sec.~\ref{sec_seeding}, where the dynamics of self-organization are studied in the presence of a laser pumping the cavity and as a function of its relative phase and amplitude. Finally, in Sec.~\ref{sec_conclusions} the conclusions are drawn.

\section{Theoretical model}\label{sec_model}

We consider a cloud of $N$ cold atoms of mass $m$ whose motion is confined along the axis of a linear standing-wave cavity as sketched in Fig.~\ref{fig_system}. The atoms are directly illuminated by a transverse laser beam whose frequency $\omegap$ is far detuned from any internal atomic transition, but close to a single cavity resonance such that the particles can scatter photons from the driving laser into this resonator mode. A second laser at the same frequency $\omegap$ directly drives the resonator mode through one of the cavity mirrors. In this limit the external and cavity degrees of freedom undergo a coupled dynamics. We denote by $x_j$ and $p_j$ the canonically-conjugated position and momentum of the $j$th atom, while $a$ and $a^\dagger$ are the bosonic annihilation and creation operators of a cavity photon. The coherent dynamics are then described by the Hamiltonian~\cite{asboth2005selforganization}
\begin{multline}\label{eq_H}
 H=\sum_{j=1}^N\left[\frac{p_j^2}{2m}+\hbar U_0 a^\dagger a\sin^2(kx_j)\right]-\hbar\DC a^\dagger a\\
 +\sum_{j=1}^N\hbar\eta\left(a+a^\dagger\right)\sin(kx_j) -i\hbar\left(\etap^*a-\etap a^\dagger\right),
\end{multline}
where $k$ denotes the cavity wave number. The parameter $U_0<0$ is the light shift per photon and $\eta\in\mathbb{R}$ is the effective cavity pump strength stemming from the light scattered from the transverse laser into the mode by the atom cloud. In addition, we also consider a longitudinal cavity pump of strength $\etap\in\mathbb{C}$, including the possibility of a different phase with respect to the transverse laser. The Hamiltonian is reported in the reference frame rotating at the transverse frequency $\omegap$, which is assumed to be resonant with the laser pumping the cavity. In particular, $\DC:=\omegap-\omegac$ is the detuning between the pump lasers and the bare cavity resonance frequency. 
\par
Owing to the non-perfect mirrors photons leak out of the resonator. These processes are taken into account in the master equation for the joint atom--field density matrix $\rho$,
\begin{equation}\label{eq_master}
 \dot\rho=\frac{1}{i\hbar}\left[H,\rho\right]+\mathcal{L}\rho,
\end{equation}
where the Liouvillean
\begin{equation}
 \mathcal{L}\rho=\kappa\left(2a\rho a^\dagger-a^\dagger a\rho-\rho a^\dagger a\right)
\end{equation}
describes cavity decay at rate $\kappa$~\cite{gardinerbook}. Dissipation effects due to atomic spontaneous emission are here neglected under the assumption that the atomic transition is driven far-off resonance. Note that the effective model in Eq.~\eqref{eq_H} is generally valid for any kind of linearly polarizable particles which can be confined within an optical resonator~\cite{ritsch2013cold}.
\par
When the thermal energy $\kB T$ of the atoms is much larger than the recoil energy $\ER\equiv\hbar\omrec:=\hbar^2k^2/2m$, i.e., $\kB T\gg\ER$, the system dynamics can be described within a semiclassical approximation and the master equation~\eqref{eq_master} can be mapped onto the following set of coupled It\=o stochastic differential equations (SDEs)~\cite{gardinerbook,domokos2001semiclassical,asboth2005selforganization},
\begin{subequations}\label{eq_sdes}
\begin{align}
 \dd x_j &= \frac{p_j}{m}\,\dd t\\
 \dd p_j &= -\frac{\partial U(x_j,\alpha)}{\partial x_j}\,\dd t\\
 \dd \alpha &= \left[i\left(\DC-U_0 \sum_{j=1}^N \sin^2(kx_j)\right)-\kappa\right]\alpha\,\dd t+\etap\,\dd t \notag\\
 &\quad-i\eta\sum_{j=1}^N \sin(kx_j)\,\dd t+\sqrt{\frac{\kappa}{2}}\left(\dd W_1+i\,\dd W_2\right),\label{eq_sde_alpha}
\end{align}
\end{subequations}
with the single-particle potential
\begin{equation}\label{eq_potential}
 U(x,\alpha)=\hbar U_0|\alpha|^2 \sin^2(kx)+\hbar\eta (\alpha+\alpha^*) \sin(kx).
\end{equation}
The term $(\dd W_1+i\,\dd W_2)/\sqrt{2}$ describes a complex Wiener process~\cite{kloedenbook} accounting for cavity input noise. In this work we numerically investigate the dynamics and steady state of this coupled system as a function of the cavity drive $\etap$.

\section{Pattern formation and seeding}\label{sec_seeding}

The coupled system of atoms and cavity photons is known to exhibit self-organized patterns in the absence of any longitudinal laser ($\etap=0$) when the transverse laser intensity exceeds a certain threshold value~\cite{domokos2002collective,black2003observation,ritsch2013cold,asboth2005selforganization}. The intracavity field is maximized when the atoms order in one of two equivalent (nonhomogeneous) configurations. The phase of the former, however, depends on the specific pattern realized, i.e., the particles either gather at lattice sites where $\sin(kx)=1$ or where $\sin(kx)=-1$ (denoted by ``even'' and ``odd'' sites, respectively).
\par
It has been argued that the occurrence of self-organization is a symmetry-breaking process, where the symmetry between the two configurations is spontaneously broken by initial particle fluctuations and cavity input noise~\cite{asboth2005selforganization}. Microscopically, one pattern is chosen for each trajectory, to which one can associate an order parameter, here identified with the quantity~\cite{asboth2005selforganization}
\begin{equation}
  \Theta:=\frac{1}{N}\sum_{j=1}^N \sin(kx_j).
\end{equation}
Below threshold, where the atoms are homogeneously distributed, $\Theta=0$, while in the perfectly self-organized phase the order parameter adopts values $\Theta=+1$~$(\Theta=-1)$ for even (odd) patterns. At this point we have to distinguish between the \emph{instantaneous} temperature-dependent critical pump strength $\eta_\mathrm{crit}$~\cite{asboth2005selforganization}, i.e., the minimal laser power required for triggering the self-organization process of a thermal gas of temperature $\kB T$, and the temperature-independent \emph{self-consistent} threshold $\etac$, above which the phase transition sets in on a long-time scale as a result of dissipation (cooling)~\cite{niedenzu2011kinetic}.
\par
\begin{figure}
  \includegraphics[width=\columnwidth]{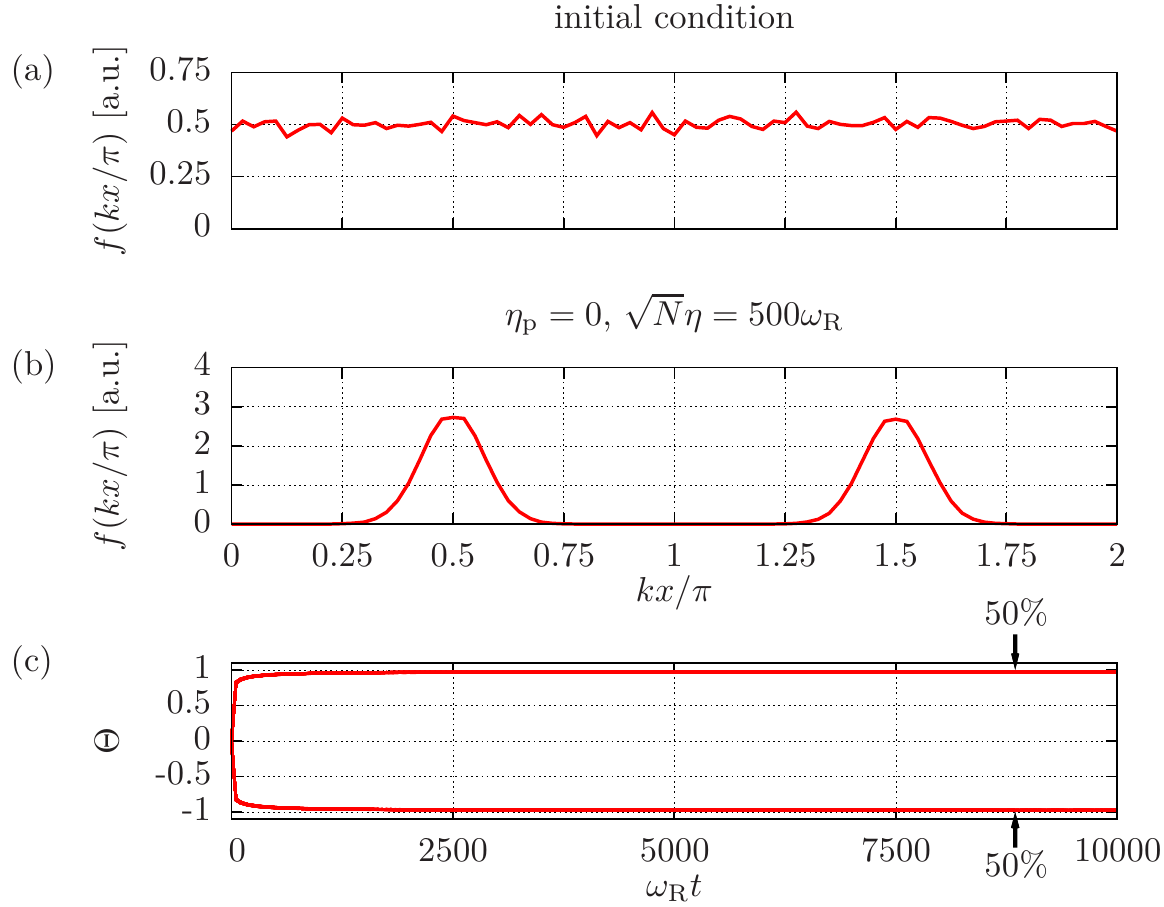}
  \centering
  \caption{(Color online) Simulation of self-organization of atoms driven by a laser and coupled to a high-finesse cavity for $\etap=0$, i.e., without additional longitudinal pump. (a) Spatial distribution at $t=0$ and (b) after reaching the steady state at $t=10^4\omrec^{-1}$. The time evolution of the order parameter $\Theta$ is displayed in (c). The curves have been obtained by numerically integrating the SDEs~\eqref{eq_sdes}. Parameters: $N=1000$, $\sqrt{N}\eta=500\omrec$, $NU_0=-100\omrec$, $\kappa=100\omrec$, and $\DC=NU_0/2-\kappa$. Ensemble average over $50\times10$ trajectories, i.e., 50 initial conditions and 10 realizations of the white noise process. The critical pump strength for the considered parameters and initial gas temperature $\kB T=2\hbar\kappa$ is $\sqrt{N}\eta_\mathrm{crit}=200\omrec$.}\label{fig_selforg}
\end{figure}

The process of self-organization is exemplified in Fig.~\ref{fig_selforg}. Figure~\ref{fig_selforg}(a) displays the initial average particle distribution when the transverse laser is switched on. The distribution after a sufficiently long time has elapsed such that the system has reached its steady state is shown in Fig.~\ref{fig_selforg}(b); one observes localization at the even and odd sites; for each trajectory only one of the two configurations is reached. Finally, Fig.~\ref{fig_selforg}(c) shows $\Theta$ as a function of time.

In summary, organization of the atoms in spatially ordered patterns corresponds to the light-induced formation of Bragg gratings. The atoms elastically scatter photons into the cavity and the intracavity field is maximized when all atoms scatter in phase, which here corresponds to arrays with interparticle distance equal to the cavity mode wavelength $\lambda=2\pi/k$. The intracavity light field gives rise to a potential which has minima at either the even or odd sites, which form from initial fluctuations with equal probability. If the cavity is directly pumped as well, the scenario drastically changes. In this case, in fact, the cavity field is the coherent superposition of the scattered and of the directly injected field as seen in Eq.~\eqref{eq_sde_alpha}. Clearly, the phase of the injected field plays a role by favoring one pattern over the other, depending on which one maximizes the depth of the intracavity potential. This allows one to ``seed'' the emergence of a specific spatial pattern above threshold, provided that the cavity drive $|\etap|$ is sufficiently strong. 

\begin{figure}
  \includegraphics[width=\columnwidth]{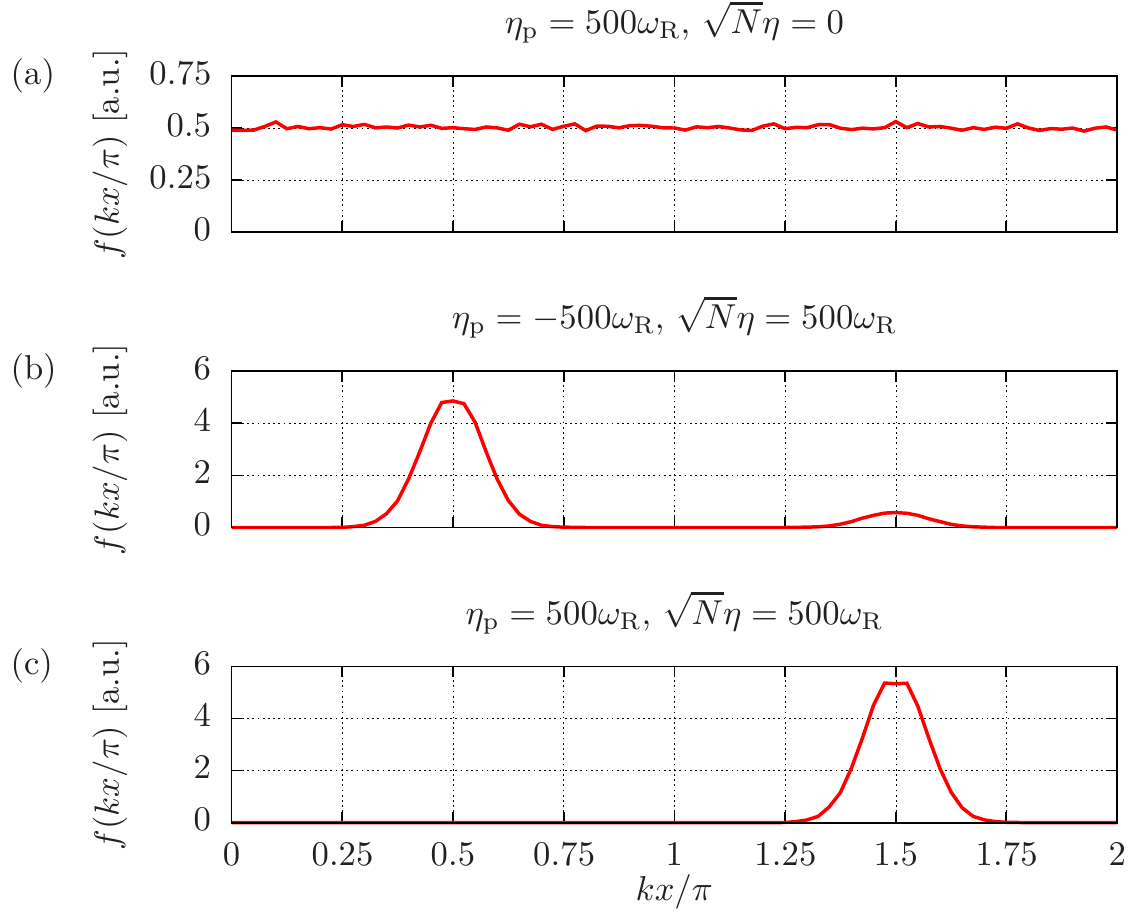}
  \centering
  \caption{(Color online) Spatial distribution of the atoms after reaching the steady state at $t=10^4\omrec^{-1}$ (a) when the transverse laser is switched off ($\eta=0$) and the cavity is pumped by a laser of intensity $\etap=500\omrec$, corresponding to a potential depth $V_0\sim 2\ER$. Subplots (b) and (c) show the spatial distribution in the presence of both longitudinal and transverse pump, with $\etap=-500\omrec$ and $\etap=500\omrec$, respectively. Ensemble average over $20\times10$ trajectories. The other parameters are the same as in Fig.~\ref{fig_selforg}.}\label{fig_seed_hist}
\end{figure}

\begin{figure}
  \includegraphics[width=\columnwidth]{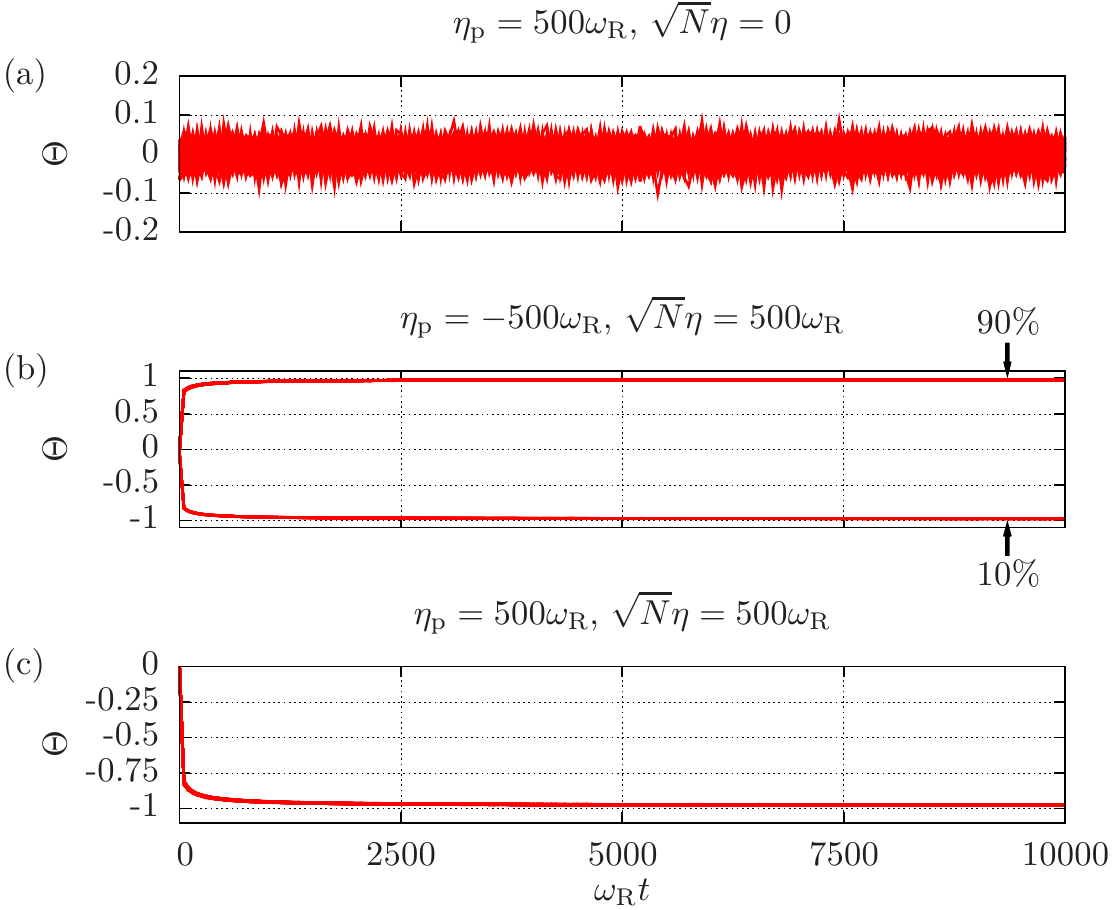}
  \centering
  \caption{(Color online) Order parameter $\Theta$ (for all trajectories) as a function of time for the parameters corresponding to the three subplots of Fig.~\ref{fig_seed_hist}, respectively.}\label{fig_seed_theta}
\end{figure}

\begin{figure}
  \includegraphics[width=\columnwidth]{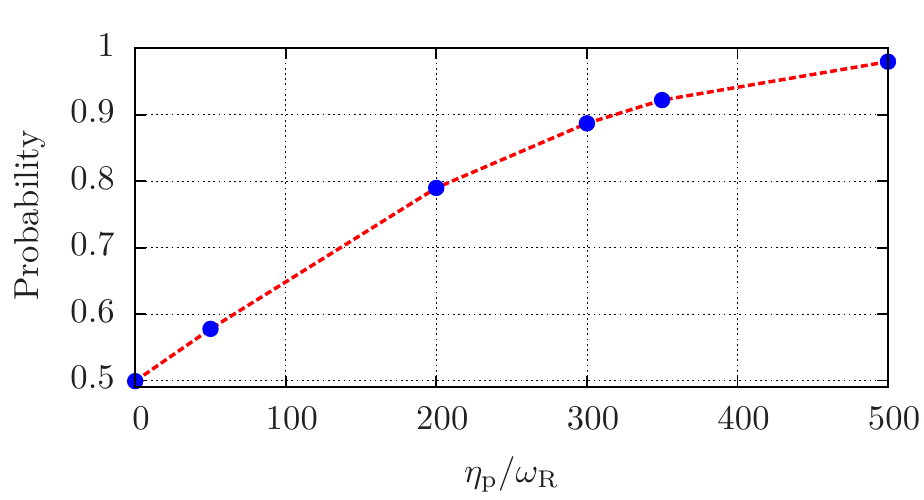}
  \centering
  \caption{(Color online) Probability that the atoms organize in an odd pattern as a function of the longitudinal pump strength $\etap\geq0$. The second pattern occurs with a small, but finite probability which decreases as $\etap$ is increased. This figure was obtained by integrating the SDEs~\eqref{eq_sdes} for a short time ($t=1\omrec^{-1}$) and computing the ratio of the number of trajectories for which $\Theta<0$ and the ensemble size. Ensemble average over $5000\times5$ trajectories. The other parameters are the same as in Fig.~\ref{fig_selforg}.}\label{fig_probability}
\end{figure}

The effect of a ``seeding'' field on the atomic spatial distribution is shown in Fig.~\ref{fig_seed_hist}. We choose the dipole potential associated with the cavity pump to be sufficiently shallow so that the atoms remain spatially uniformly distributed when the transverse laser is off; \cf Fig.~\ref{fig_seed_hist}(a). Starting from this situation, a transverse laser (atom pump) is switched on at time $t=0$. Figures~\ref{fig_seed_hist}(b) and~\ref{fig_seed_hist}(c) display the two configurations which are obtained after a transient time when the relative phase between the two lasers is set either equal to 0 or $\pi$. We observe that the symmetry between the two patterns is broken compared to Fig.~\ref{fig_selforg}(b); \cf also the corresponding order parameter in Fig.~\ref{fig_seed_theta}. The atoms are with a high probability either localized in the even or in the odd sites, respectively, depending on the phase difference between the two driving lasers. We see though in Figs.~\ref{fig_seed_hist}(b) and~\ref{fig_seed_theta}(b) that the second pattern is only strongly suppressed and not impossible; there roughly 10\% of the trajectories ended up in the ``wrong'' configuration. This specific number is an artifact of the relatively small ensemble considered, just like the fact that all trajectories ended up in an odd pattern for positive $\etap$ in Figs.~\ref{fig_seed_hist}(c) and~\ref{fig_seed_theta}(c). These two cases thus already suggest a large statistical error of the seeding efficiency and the necessity of averaging over much larger ensembles. As can be seen in Fig.~\ref{fig_probability} for a large ensemble the probability that a certain configuration is dynamically realized approaches 100\% for sufficiently large values of the cavity drive $|\etap|$.

\subsection{Stationary patterns}

Pattern seeding by means of a longitudinal field can be understood in terms of interference between the electric fields scattered by the atoms and directly injected by the cavity pump. This determines the intracavity field amplitude and therefore the depth of the $\sin(kx)$ component of the potential~\eqref{eq_potential}, which is proportional to the field's real part. Neglecting noise, the steady-state solution of Eq.~\eqref{eq_sde_alpha} for the intracavity field reads
\begin{equation}\label{eq_alpha_ss}
 \alpha_\mathrm{ss}=\frac{-i\eta N\Theta+\etap}{\kappa-i\Delta},
\end{equation}
with the effective detuning $\Delta:=\DC-NU_0\mathcal{B}<0$ and the bunching parameter $\mathcal B:=\frac{1}{N}\sum_j \sin^2(kx_j)$~\cite{asboth2005selforganization}. This result is valid for $kv_\mathrm{T}\ll\kappa$, where $v_\mathrm{T}:=\sqrt{2\kB T/m}$ is the thermal velocity. Let us assume for simplicity that also $\etap\in\mathbb{R}$. Then, for a given sign of $\Theta$---i.e., whether the atoms form an even or odd pattern---the sign of $\etap$ determines whether $|\realt{\alpha}|$ becomes larger and hence the potential deeper. For instance, for $\etap>0$ we expect that odd patterns ($\Theta<0$) will be energetically favored and therefore occur with larger probability as visible from Fig.~\ref{fig_probability}.

\begin{figure}
 \centering
 \includegraphics[width=\columnwidth]{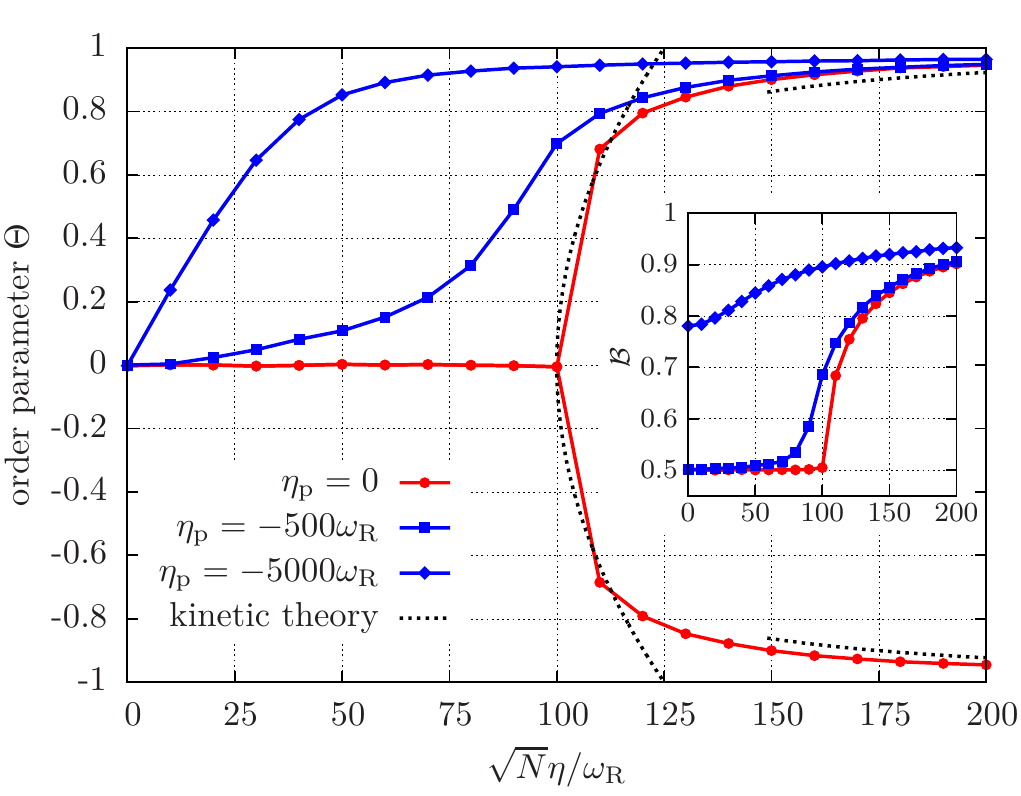}
 \caption{(Color online) Order parameter $\Theta$ evaluated in steady state (at $t=20N \omrec^{-1}$) for $\etap=0$ (red dots), $\etap=-500\omrec$ (blue squares), and $\etap=-5000\omrec$ (blue diamonds). For $\etap<0$ the order parameter is always positive when driving the cavity. The black-dotted lines are the asymptotic steady-state predictions from kinetic theory in the weak-coupling limit~\cite{niedenzu2011kinetic}. Inset: corresponding bunching parameter $\mathcal{B}$. Ensemble average of $5\times 5$ trajectories. The other parameters are the same as in Fig.~\ref{fig_selforg}.}\label{fig_phasediagram_long}
\end{figure}

We now analyze the steady-state order parameter $\Theta$ as a function of the atom pump amplitude $\eta$ for a chosen value of the phase and magnitude of the cavity pump $\etap$. Figure~\ref{fig_phasediagram_long} displays this quantity for $\etap\leq 0$ as obtained by numerically integrating Eqs.~\eqref{eq_sdes} for long times. For completeness we also report the analytical predictions of~\cite{niedenzu2011kinetic} valid at $\etap=0$ in the weak-coupling regime ($N|U_0|\ll|\DC|$). Indeed, whilst we observe the expected bifurcation at the self-consistent threshold without cavity drive, only one branch is selected when the latter is sufficiently strong. For larger ensembles, however, the behavior suggested by Fig.~\ref{fig_probability} is expected to become visible, i.e., a finite number of trajectories ending up in the opposite pattern. In particular, finite probabilities of finding odd patterns are expected at larger values of $\eta$. At the same time the sharp transition is smeared out to a smooth crossover---the value of $\Theta$ increases monotonously to unity. This behavior becomes more enhanced as $|\etap|$ is increased by one order of magnitude.

A special case is realized when the longitudinal pump is sufficiently strong to give rise to a deep intracavity lattice even in the absence of the transverse laser. This is found, for instance, when $\etap=-5000\omrec$. In this regime, at $\eta=0$ the atoms are localized at the minima of the $\sin^2(kx)$ part of the cavity optical lattice~\eqref{eq_potential} which is reflected by a high value of the bunching parameter $\mathcal{B}$, as visible in the inset of Fig.~\ref{fig_phasediagram_long}. A small value of $\eta$ then already gives rise to a finite probability of finding the atoms in the even pattern. We have checked that the observed steady-state value of $\Theta$ is mainly due to the mechanical potential associated with the longitudinal laser. For $\etap=-500\omrec$, on the other hand, collective photon scattering plays a crucial role in the formation of even patterns.

\subsection{Dynamical build-up of the organized phase}

\begin{figure}
 \centering
 \includegraphics[width=\columnwidth]{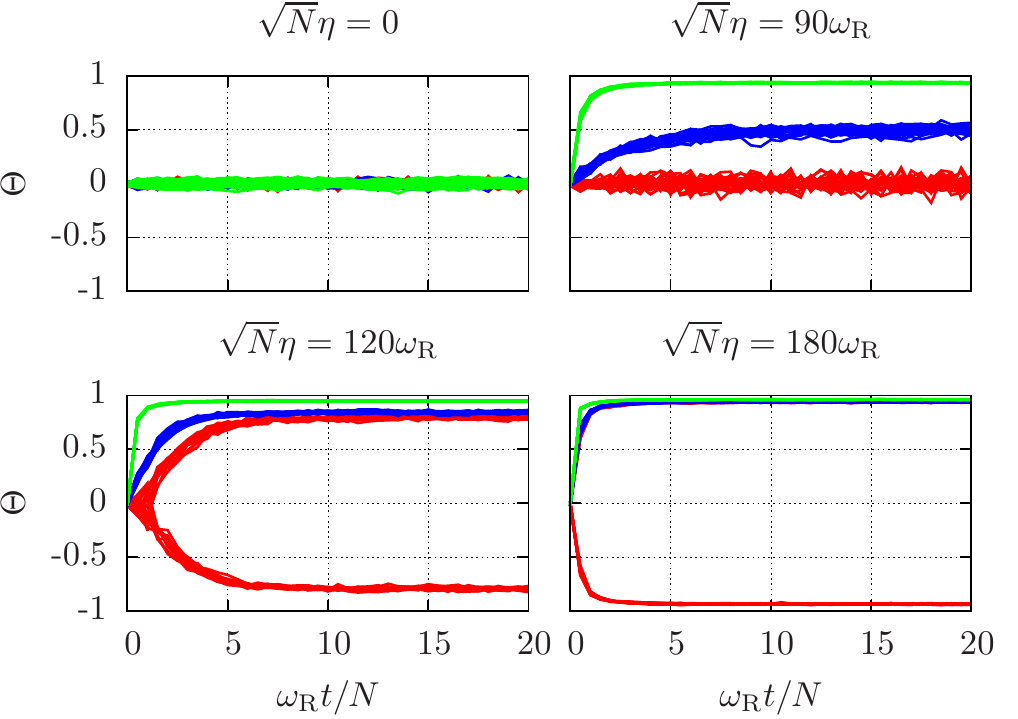}
 \caption{(Color online) Order parameter $\Theta$ for $5\times5$ trajectories simulated using Eqs.~\eqref{eq_sdes} as a function of time for $\etap=0$ (red), $\etap=-500\omrec$ (blue), and $\etap=-5000\omrec$ (green). From top to bottom (left to right) the value of $\eta$ is $\sqrt{N}\eta=(0,90,120,180)\omrec$. The other parameters are the same as as in Fig.~\ref{fig_selforg}.}\label{fig_phasediagram_dynamical}
\end{figure}

We now analyze the formation of a seeded pattern (e.g., the even one) considering two situations. First, when the atoms' initial spatial distribution is uniform, and second when the initial distribution corresponds to the opposite pattern. Figure~\ref{fig_phasediagram_dynamical} shows the onset of a seeded pattern for an initially uniform distribution, i.e., how the order parameter in Fig.~\eqref{fig_phasediagram_long} is dynamically established. We observe that the time scale over which the pattern forms decreases as the amplitude of the seeding field is increased. This behavior is particularly pronounced below threshold.

Let us now assume that $\etap=0$ and that the atoms are pumped by a transverse laser with $\eta>0$ above the (instantaneous) threshold. After some time the system reaches a stationary configuration which can be, for instance, an even pattern. An intense longitudinal pump field is then switched on. If its amplitude $\etap$ is chosen to be real, the particles remain in the even pattern since according to Eq.~\eqref{eq_alpha_ss} the number of intracavity photons---and hence the potential depth---increases. Instead, when for instance the relative phase of the longitudinal pump differs from the transverse laser field by $\pi/2$, then the scattered and the injected field can interfere. This situation can lead to a pattern flip in the case that the phase of the scattered field gives rise to destructive interference. We expect that the scattered field vanishes and thus the atoms reorganize in the pattern for which the two field contributions add up coherently. We thus choose the relative phase $\pi/2$ and $|\etap|$ about 40 times larger than the maximum scattering rate into the resonator by the atoms ($\sqrt{N}\eta$). 

\begin{figure}
 \centering
 \includegraphics[width=\columnwidth]{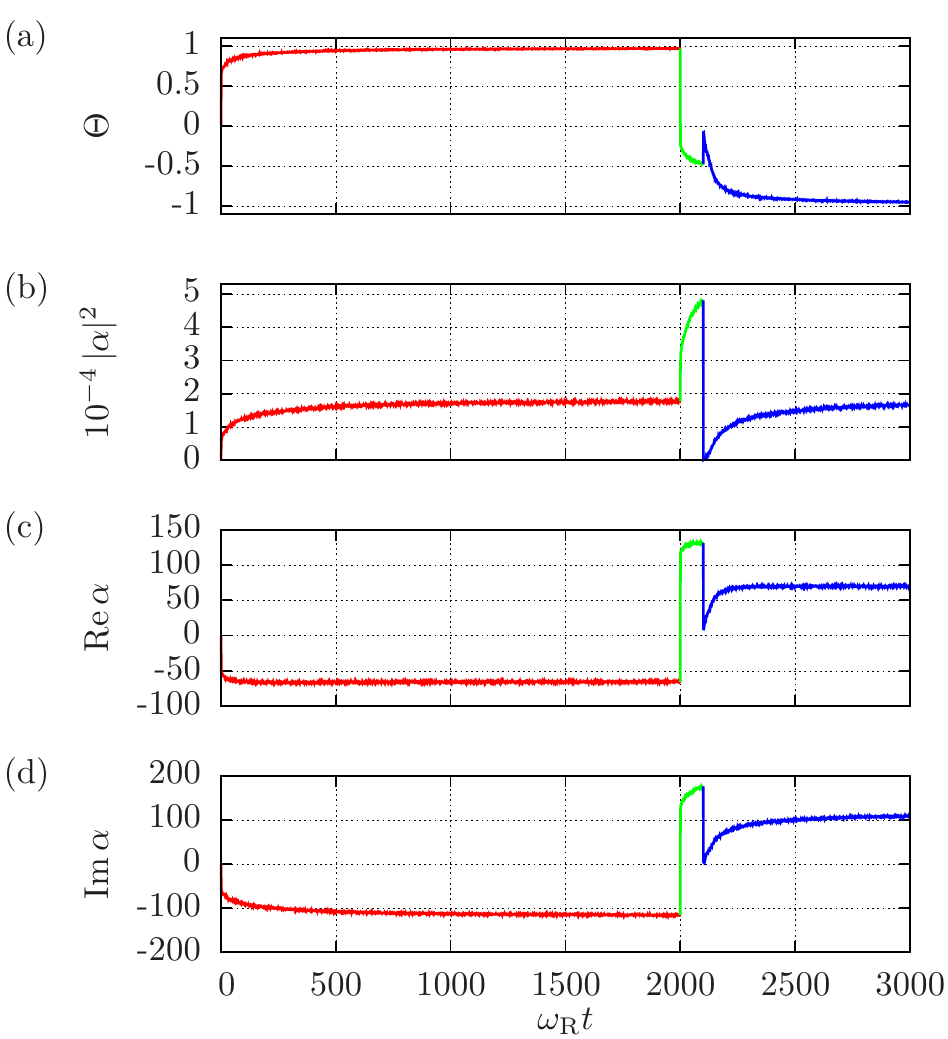}
 \caption{(Color online) Example of dynamically flipping patterns with a longitudinal pump. The system behavior is shown as a function of time for a sequential change of the cavity drive by plotting (a) the order parameter, (b) the intracavity photon number, and the real (c) and imaginary (d) part of the field amplitude. First, $\etap=0$ and $\sqrt{N}\eta=500\omrec$. At $t=2000\omrec^{-1}$ the longitudinal laser is switched on with $\etap=2i\times 10^{4}\omrec$ in order to compensate for the scattered field for a short time. Finally, at $t=2100\omrec^{-1}$ the external laser is reduced to $\etap=500\omrec$. The other parameters are the same as in Fig.~\ref{fig_selforg}.}\label{fig_flip}
\end{figure}

Figure~\ref{fig_flip} shows the order parameter, mean number of intracavity photons, and the real and imaginary part of the intracavity field amplitude as a function of time for a single trajectory. When the strong longitudinal pump is switched on, one observes that the system readjusts in an odd pattern which augments the intracavity photon number. The order parameter, however, remains smaller than unity, representing a situation in which several defects, namely atoms trapped at even sites, are present. This can be understood through the SDE~\eqref{eq_sde_alpha} for the field and its approximate steady-state value~\eqref{eq_alpha_ss}: the intracavity field is overcompensated by the external driving such that its real part becomes positive, i.e., the odd sites become deeper than the even ones according to Eq.~\eqref{eq_potential}. Due to the external pump the latter, however, remain sufficiently deep to confine a considerable fraction of the particles on the considered short time scale. Afterwards, the longitudinal laser intensity is reduced to $\etap=500\omrec$, for which the intracavity optical lattice is shallow, and the particles are now again trapped by their own scattered light, i.e., after a transient time, in which the atoms first form a uniform distribution, the order parameter becomes again close to unity. Its sign, however, is the opposite of the initial one prior to the pulse sequence. Alternatively, one could also envisage a scheme, where the longitudinal pump power is continuously reduced after the flip. This behavior is exemplary for understanding how the system reacts to external perturbations. 
 
\section{Conclusions and outlook}\label{sec_conclusions}

In this work we presented a model and a numerical study of the effect of a longitudinal pump on self-organization of cold atoms in an optical resonator. We have focused on the situation in which both the transverse laser, pumping the atoms, and the longitudinal laser, driving the cavity mode, are resonant. The system exhibits a stationary state which can be an ordered pattern even below the self-organization threshold, provided that the longitudinal pump is sufficiently strong. The phase of the pump is crucial in determining the pattern which self-organizes. The longitudinal field, hence, acts as a seed, breaking the symmetry between the even and odd patterns, which are otherwise equivalent. The relative phase between longitudinal and transverse pump is thus a control handle for determining the configuration in which the atoms self-organize, to the point that it can be used to force the atoms to flip patterns. 
\par
Our analysis has so far focused on resonant external fields, which allow one to reduce the dynamics to a time-independent problem by moving to the reference frame rotating at the laser frequency. The situation is going to be dramatically modified when the two lasers are detuned one from the other. In this case the equations of motion are explicitly time dependent and exhibit an intrinsic period determined by the frequency mismatch between the two lasers. Chaotic behavior could here emerge at sufficiently low levels of noise. 
\par
An extension to the multimode case, e.g., ring cavities---where a continuous symmetry is broken---might also be interesting.

\begin{acknowledgments}
The authors acknowledge discussions with Tobias Donner and Andr\'{a}s Vukics. W.N.\ thanks the Universit\"at des Saarlandes, where parts of this work were performed, for hospitality. This work has been supported by the Austrian Science Fund FWF through project F4013, the EU (IP AQUTE, STREP PICC), the German BMBF (QuORep), and the German Research Foundation (DFG).
\end{acknowledgments}

\end{document}